\documentclass[11pt,longbibliography,titlepage,showkeys]{revtex4-1}
\usepackage{amsfonts,amssymb,amsmath,bm}

\begin{document}

\def\diag{\mathop{\rm diag}\nolimits}
\def\sh{\mathop{\rm sh}\nolimits}
\def\ch{\mathop{\rm ch}\nolimits}
\def\var{\mathop{\rm var}}\def\exp{\mathop{\rm exp}\nolimits}
\def\Re{\mathop{\rm Re}\nolimits}
\def\Sp{\mathop{\rm Sp}\nolimits}
\def\kp{\mathop{\text{\ae}}\nolimits}
\def\bk{{\bf {k}}}
\def\bp{{\bf {p}}}
\def\bq{{\bf {q}}}
\def\lra{\mathop{\longrightarrow}}
\def\Const{\mathop{\rm Const}\nolimits}
\def\sh{\mathop{\rm sh}\nolimits}
\def\ch{\mathop{\rm ch}\nolimits}
\def\var{\mathop{\rm var}}
\def\mK{\mathop{{\mathfrak {K}}}\nolimits}
\def\mR{\mathop{{\mathfrak {R}}}\nolimits}
\def\mv{\mathop{{\mathfrak {v}}}\nolimits}
\def\mV{\mathop{{\mathfrak {V}}}\nolimits}
\def\mD{\mathop{{\mathfrak {D}}}\nolimits}
\def\mN{\mathop{{\mathfrak {N}}}\nolimits}
\def\mS{\mathop{{\mathfrak {S}}}\nolimits}

\newcommand\ve[1]{{\mathbf{1}}}

\def\Re{\mbox {Re}}
\newcommand{\Z}{\mathbb{Z}}
\newcommand{\R}{\mathbb{R}}
\def\mg{\mathop{{\mathfrak {g}}}\nolimits}
\def\mK{\mathop{{\mathfrak {K}}}\nolimits}
\def\mk{\mathop{{\mathfrak {k}}}\nolimits}
\def\mR{\mathop{{\mathfrak {R}}}\nolimits}
\def\mv{\mathop{{\mathfrak {v}}}\nolimits}
\def\mV{\mathop{{\mathfrak {V}}}\nolimits}
\def\mD{\mathop{{\mathfrak {D}}}\nolimits}
\def\mN{\mathop{{\mathfrak {N}}}\nolimits}
\def\ml{\mathop{{\mathfrak {l}}}\nolimits}
\def\mf{\mathop{{\mathfrak {f}}}\nolimits}
\newcommand{\ccm}{{\cal M}}
\newcommand{\cE}{{\cal E}}
\newcommand{\cV}{{\cal V}}
\newcommand{\cI}{{\cal I}}
\newcommand{\cR}{{\cal R}}
\newcommand{\cK}{{\cal K}}
\newcommand{\cH}{{\cal H}}
\newcommand{\cW}{{\cal W}}

\def\div{\mathop{\rm div}\nolimits}
\def\br{\mathop{{\bf {r}}}\nolimits}
\def\bS{\mathop{{\bf {S}}}\nolimits}
\def\bA{\mathop{{\bf {A}}}\nolimits}
\def\bJ{\mathop{{\bf {J}}}\nolimits}
\def\bn{\mathop{{\bf {n}}}\nolimits}
\def\bg{\mathop{{\bf {g}}}\nolimits}
\def\bv{\mathop{{\bf {v}}}\nolimits}
\def\be{\mathop{{\bf {e}}}\nolimits}
\def\bp{\mathop{{\bf {p}}}\nolimits}
\def\bz{\mathop{{\bf {z}}}\nolimits}
\def\bbf{\mathop{{\bf {f}}}\nolimits}
\def\bb{\mathop{{\bf {b}}}\nolimits}
\def\ba{\mathop{{\bf {a}}}\nolimits}
\def\bx{\mathop{{\bf {x}}}\nolimits}
\def\by{\mathop{{\bf {y}}}\nolimits}
\def\br{\mathop{{\bf {r}}}\nolimits}
\def\bs{\mathop{{\bf {s}}}\nolimits}
\def\bH{\mathop{{\bf {H}}}\nolimits}
\def\bk{\mathop{{\bf {k}}}\nolimits}
\def\be{\mathop{{\bf {e}}}\nolimits}
\def\bnul{\mathop{{\bf {0}}}\nolimits}
\def\bq{{\bf {q}}}

\newcommand{\oV}{\overline{V}}
\newcommand{\vkp}{\varkappa}
\newcommand{\os}{\overline{s}}
\newcommand{\opsi}{\overline{\psi}}
\newcommand{\ov}{\overline{v}}
\newcommand{\oW}{\overline{W}}
\newcommand{\oPhi}{\overline{\Phi}}

\def\mI{\mathop{{\mathfrak {I}}}\nolimits}
\def\mA{\mathop{{\mathfrak {A}}}\nolimits}

\def\st{\mathop{\rm st}\nolimits}
\def\tr{\mathop{\rm tr}\nolimits}
\def\sign{\mathop{\rm sign}\nolimits}
\def\d{\mathrm{\rm d}}
\def\const{\mathop{\rm const}\nolimits}
\def\O{\mathop{\rm O}\nolimits}
\def\Spin{\mathop{\rm Spin}\nolimits}
\def\exp{\mathop{\rm exp}\nolimits}
\def\SU{\mathop{\rm SU}\nolimits}
\def\mU{\mathop{{\mathfrak {U}}}\nolimits}
\newcommand{\cU}{{\cal U}}
\newcommand{\cD}{{\cal D}}

\def\mI{\mathop{{\mathfrak {I}}}\nolimits}
\def\mA{\mathop{{\mathfrak {A}}}\nolimits}
\def\mU{\mathop{{\mathfrak {U}}}\nolimits}

\def\st{\mathop{\rm st}\nolimits}
\def\tr{\mathop{\rm tr}\nolimits}
\def\sign{\mathop{\rm sign}\nolimits}
\def\const{\mathop{\rm const}\nolimits}
\def\O{\mathop{\rm O}\nolimits}
\def\Spin{\mathop{\rm Spin}\nolimits}
\def\exp{\mathop{\rm exp}\nolimits}

\title{General relativity  and precision tests of fundamental symmetries}

\author{N.N. Nikolaev$^*$ and S.N. Vergeles$^{**}$}

\address{Landau Institute for Theoretical Physics,
Russian Academy of Sciences,
Chernogolovka, Moscow region, 142432 Russia\\
$^*$E-mail: nikolaev@itp.ac.ru\\
$^{**}$E-mail: vergeles@itp.ac.ru}

\begin{abstract}


Search for the Electric Dipole Moment of nuclear particles is at the forefront of incessant quest for CP violation beyond Standard Model.
The ultimate target is to reach a sensitivity to the electric dipole moment of neutrons, protons, deuterons etc. at the level of
$\sim 10^{-15}$ nuclear magnetons. Defying the common lore on weakness of gravity, spurious signals induced by
curved space-time in the gravity field of the rotating Earth become quite substantial at such a daunting sensitivity. We review the recent development in the field with an  emphasis on the geometric magnetic field in pure electrostatic systems at rest on the rotating Earth.

\end{abstract}


\maketitle

\section{INTRODUCTION}\label{aba:sec1}

A subject of this review is an impact of General Relativity (GR) on precision tests of discrete fundamental symmetries in atomic, nuclear and particle physics.
The interest in observation of the P- and T(CP)-noninvariant electric dipole moment (EDM)  of charged particles in
storage rings is fuelled by a potential to resolve the outstanding failure of the Standard Model (SM), which by 9 orders in magnitude is much too feeble to
explain the observed baryon asymmetry of the Universe \cite{sakharov1991pis,bernreuther2002cp,chupp2019electric}.

Experimental signal of EDM is the spin precession in the electric field. If one parametrizes the EDM of nucleons and
light nuclei, $d=\eta_{EDM}\mu_N$, in units of the nuclear magneton
$\mu_N$, then widely discussed models of CP violation suggest $\eta_{EDM} \sim 10^{-10}$, {\it i.e.}, $d\sim 10^{-24}$ e$\cdot$cm,
whereas the Standard Model predicts $\eta_{EDM} \sim 10^{-17}$ and  $d\sim 10^{-31}$ e$\cdot$cm $ \ $ \cite{okun1967violation,
khriplovich2012cp,chupp2019electric}.

Spins of charged particles can be subjected to electric fields only in storage rings. Experiments with protons in storage rings can in principle achieve the
sensitivity to  $\eta_{EDM} \sim 10^{-15} \ \ $ {\it i.e.,} to the EDM $d_p \sim 10^{-29}\, e\cdot$cm, corresponding to the angular velocity of the EDM-induced spin rotation $ \Omega_{EDM} \sim  10^{-9} rad/s \ \ $ \cite{anastassopoulos2016storage, rathmann2019electric,abusaif2019storage}.
To experimentally  disentangle such a minuscule effects, it is imperative to understand the spin dynamics and to control background effects at the same level.

The central issue is that our terrestrial laboratories reside in the curved space-time of rotating Earth and are subject to non-inertial motion. Silenko and Teryaev were the first to discuss the impact of Earth's gravity pull on spin dynamics in storage rings \cite{Silenko}. In  all magnetic rings the found effect is of pure academic interest.  The issue becomes acute in the all-electric frozen-spin storage rings, considered the Holy Grail to search for the proton EDM \cite{anastassopoulos2016storage}. Here, choosing the  magic energy, one eliminates the false rotation of spin from coupling of the magnetic dipole moment (MDM) to the motional magnetic fields, and the sole source of  spin rotation is the EDM coupling to the confining radial electric field. However, the Earth's gravity pull has to be compensated for by the vertical focusing electric field, which entails the motional radial magnetic field and false spin rotation due to the MDM \cite{Orlov}. Incidentally, a commensurate geodetic spin rotation in the gravitational field proper has been  predicted more than a century ago by de Sitter \cite{deSitter,pomeranskiiUFN} and has been confirmed experimentally in the Gravity-B experiment on precession of satellite based classical gyroscope \cite{GravityProbePRL}.

Weak though the gravity is, at  $\eta_{\rm EDM} \sim 10^{-15} $ the gravity effect overtakes the EDM signal by more than one order in magnitude \cite{Orlov,Obukhov,nikolaev2019gravity}.
Fortunately, the all electric storage rings can be run with the counter-rotating beams and the T-violating EDM signal can be separated from the T-invariant gravity effect. Furthermore, the uniquely predictable gravity signal is viewed as a unique Standard Candle for calibration of the performance of all electric storage rings (see \cite{abusaif2019storage}, Appendix D).

The Orlov et al. gravity pull effect \cite{Orlov} is an important example of impact of GR on the spin dynamics in the external pure electric field. A much more subtle  twist of the same story is emergence of magnetic fields in a system which is rigorously electrostatic on the surface of rotating Earth. Indeed, to an observer residing on distant stars, the motion of static charges on the rotating Earth do obviously generate magnetic fields. Will there be any trace of the residual magnetic field to an observer comoving with charges which are static in the rotating laboratory? A thorough analysis of solution of Maxwell equations in the curved space-time of rotating bodies has revealed an existence of nontrivial geometric magnetic fields (GMF) \cite{VergelesJETP,VergelesJHEP}. A remarkable  feature of the GMF is that it can not be screened away by magnetic shielding. Principal point is that exceedingly small GR effects are still gigantic on the scale of $\eta_{\rm EDM}$ of practical interest. For instance, at EDMs such that $\eta_{\rm EDM} \sim 10^{-15}$, the angular velocity of Earth's rotation is five orders of magnitude higher than the angular velocity of the EDM-induced spin rotation. Rotation of an all electric
storage ring with opposite charged electrodes generates opposite running currents. The emerging GMF on the beam orbit  will be about $\eta_{\omega } =v_{\omega}/c \sim 10^{-11}$
times the confining electrostatic field on the orbit, where $v_{\omega}$ is the storage ring rotation velocity.

The GMFs exhibit subtle dependence on the geometry of static charge distributions. For instance, in search for the EDM of ultracold neutrons the principal false EDM signal comes from the gradient of the geometric field. Numerically, it will be sizeable  in the neutron EDM experiments of next generation. The geometric magnetic properties of the charged spherical shell closely resemble those of the magnetic dipole. In all electric storage rings the GMF lies in the ring plane and locally is a very strong background to the EDM signal. However, the specific angular dependence of the GMF nullifies the GMF driven integrated spin rotation per turn. There remains an open issue whether GMF in conjunction with orbit distortions might produce a nonvanishing geometric phase or not.

As outlined above, the two major themes of this review are the GR corrections to the motion of charged and spinning particles and  GMFs in electrostatic systems in non-inertial motion. In Section 2 we discuss the GR results for the dynamics of spinning charged particles in external electromagnetic fields.  Section 3 is devoted to derivation of the GMF field in pure electrostatic systems in terrestrial laboratories residing on the rotating Earth. A background from the GMF in the EDM experiments is treated in section 4. In the Conclusions we overview the main results and comment on open issues.  This review is bound to heavy use of the solid background of Einstein's General Relativity (GR). The curved space-time of rotating Earth is described by the Kerr metric, which unifies the Earth gravity pull and  rotation effects. Correspondingly, the review is supplemented by quite an extensive GR technicalities of relevance to spin dynamics.

\section{GR and CHARGED PARTICLE $\&$ SPIN DYNAMICS in EXTERNAL FIELDS }

\subsection{ Maxwell  equations}

Let the holonomic electromagnetic field (2-form) be expressed through 4-potential $A_{\mu}$ (1-form) in local coordinates $x^{\mu},\,\mu=0,1,2,3=(0,i)$, $i=1,2,3$ as:
\begin{equation}
F_{\mu\nu}=\partial_{\mu}A_{\nu}-\partial_{\nu}A_{\mu}.
\label{M10}
\end{equation}
Then the homogeneous Maxwell equations are satisfied automatically:
\begin{equation}
\varepsilon_{\mu\nu\lambda\rho}\partial_{\nu}F_{\lambda\rho}=0
\label{M20}
\end{equation}

The inhomogeneous Maxwell equations in the local coordinates read
\begin{eqnarray}
\frac{1}{\sqrt{-g}}\partial_{\nu}\left(\sqrt{-g}F^{\nu\mu}\right)=4\pi J^{\mu},
\nonumber \\
g\equiv\det g_{\mu\nu},  \quad F^{\mu\nu}=g^{\mu\lambda}g^{\nu\rho}F_{\lambda\rho}.
\label{M30}
\end{eqnarray}

The electric and magnetic fields in an orthonormal basis
(ONB) are defined by the usual rules (see Eq. (\ref{G64}) in Appendix A):
\begin{equation}
F_{ab}=\tilde{e}_a^{\mu}\tilde{e}_b^{\nu}F_{\mu\nu}.
\label{M35}
\end{equation}
In particular, electric and magnetic fields in ONB are defined as follows:
\begin{equation}
{\bf E}^{\alpha}=-F_{\alpha 0}=F^{\alpha 0}=e^{\alpha}_{\mu}e^0_{\nu}F^{\mu\nu}=e^{\alpha}_ie^0_0F^{i0}+e^{\alpha}_ie^0_jF^{ij},
\label{M40}
\end{equation}
\begin{equation}
\varepsilon_{\alpha\beta\gamma}{\bf H}^{\gamma}=-F_{\alpha\beta}=-F^{\alpha\beta}=-e^{\alpha}_{\mu}e^{\beta}_{\nu}F^{\mu\nu},
\quad \varepsilon_{123}=1.
\label{M50}
\end{equation}

\subsection{The charged particle momentum dynamics}


Let $\d s\equiv \sqrt{g_{\mu\nu}\d x^{\mu}\d x^{\nu}}$ be along the world line of a moving particle.  We use the standard  definition of the covariant derivative, $D X^{\mu}/\d s\equiv u^{\nu}\nabla_{\nu}X^{\mu}$,
where $u^{\nu}\equiv \d x^{\nu}/\d s$, so that $g_{\mu\nu}u^{\mu}u^{\nu}=1$. According to
Eqs. (\ref{G65}),  (\ref{G70}) in Appendix A, in terms of connections
\begin{equation}
\frac{D X^a}{\d s}=\frac{\d X^a}{\d s}+\left(\gamma^a_{b\mu}\frac{\d x^{\mu}}{\d s}\right)X^b=\frac{\d X^a}{\d s}+
\left(\gamma^a_{bc}u^c\right)X^b\, .
\label{D5}
\end{equation}
In the local coordinates, the equation of motion of a charged particle is
\begin{equation}
mc\frac{D u^{\mu}}{\d s}=\frac{q}{c}F^{\mu\nu}u_{\nu}.
\label{D10}
\end{equation}

It is instructive to look at this equation inside the "freely falling lift", in the  Riemann normal coordinates (see Eq. (\ref{G160}) in Appendix D), where one recovers the familiar results in the Cartesian coordinates:
\begin{gather}
\frac{\d \gamma}{\d t}=\frac{q}{mc}{\bf E}{\boldsymbol\beta},\quad\quad
\frac{\d(\gamma{\boldsymbol\beta}^{\alpha})}{\d t}=
\frac{q}{mc}
\left({\bf E}+[{\boldsymbol\beta}\times{\bf H}]\right)^{\alpha},
\nonumber \\
u^a=\gamma(1,\,{\boldsymbol\beta}),  \quad {\boldsymbol\beta}\equiv{\bf v}/c, \quad
\d s=\frac{c}{\gamma}\d t.
\label{D20N}
\end{gather}
Here $\d t$ is the proper time interval in the laboratory frame of reference in which the particle moves with
the velocity ${\bf v}$.  In the general case
\begin{equation}
u^a=e^a_{\mu}\frac{\d x^{\mu}}{\d s}.
\label{D22N}
\end{equation}

The fields ${\bf E}$ and ${\bf H}$ in different ONBs differ only by the Lorentz transformation. In an arbitrary ONB,  the GR effects enter via connections:
\begin{gather}
\frac{\d \gamma}{\d t}=\frac{q}{mc}{\bf E}{\boldsymbol\beta}+c(\gamma_{\alpha 0c}u^c){\boldsymbol\beta}^{\alpha},
\nonumber \\
\frac{\d(\gamma{\boldsymbol\beta}^{\alpha})}{\d t}=
\frac{q}{mc}
\left({\bf E}^{\alpha}+[{\boldsymbol\beta}\times{\bf H}]^{\alpha}\right)+
c\Big((\gamma_{\alpha 0c}u^c)+(\gamma_{\alpha\beta c}u^c){\boldsymbol\beta}^{\beta}\Big).
\label{D20}
\end{gather}
The principal point is
that the fields ${\bf E}$ and ${\bf H}$  defined in ONB according to
(\ref{M40}) and (\ref{M50}), possess all the dynamic properties of the electric and magnetic fields, correspondingly.


The angular velocity of the cyclotron rotation of the particle is
\begin{equation}
{\boldsymbol\Omega}_c=\frac{1}{{\bf v}^2}\left[{\bf v}\times\frac{\d{\bf v}}{\d t}\right] = {\boldsymbol\Omega}_c^{EM} +{\boldsymbol\Omega}_c^{GR}\,,
\label{Cycl10}
\end{equation}
where the electromagnetic (EM) and GR components  in the laboratory frame
equal
\begin{equation}
{\boldsymbol\Omega}_c^{EM}=-\frac{q}{mc\gamma}{\bf H}+\frac{q}{mc\gamma{\boldsymbol\beta}^2}\bigg(({\bf H}{\boldsymbol\beta}){\boldsymbol\beta}+[{\boldsymbol\beta}\times{\bf E}]
\bigg),
\label{Cycl20}
\end{equation}
\begin{gather}
{\boldsymbol\Omega}^{GR}_c=\frac{1}{\gamma{\bf v}^2}{\Huge\Big[}{\bf v}\times \,\,\Bigg\{
\left(\frac{2\gamma^2-1}{\gamma}{\boldsymbol g}_0+\gamma\big[{\boldsymbol\omega}\times [{\bf R}\times {\boldsymbol\omega}]\big]\right)
\nonumber \\
+2\gamma\left(\left(1-\frac{r_g}{2R_{\oplus}}(1-I)\right)+\frac{3}{4c^2} \big({\bf \omega}[{\boldsymbol R} \times {\bf v}]\big)\right)[{\bf v}\times {\boldsymbol\omega}]
\nonumber \\
+\frac{2\gamma r_g}{R_{\oplus}^2}\left(1-\frac32I\right)
({\boldsymbol\omega}{\bf n}_{\oplus})[{\bf R}\times {\bf v}]\Bigg\}{\Huge\Big]},
\label{Cycl30}
\end{gather}
respectively.
 Hereafter we focus on the terrestrial laboratory, {\it i.e.,}  $|{\bf R}|=R_{\oplus}$, ${\bf n}_{\oplus}={\bf R}/R_{\oplus}$,
${\bf g}_0/c^2\equiv-(r_g/2R_{\oplus}^2){\bf n}_{\oplus}$ (see Appendix B).

The effect of the Earth gravity pull, $\propto {\bf g}_0$, enters with the relativistic factor $(2\gamma^2-1)/g\ $ \cite{Obukhov}. The accompanying centripetal acceleration comes with the relativistic factor $\gamma$. The second term describes the familiar Coriolis acceleration with the relativistic factor  $\gamma$ and the GR correction factor, in which the term with the vector product $[{\bf R}\times {\bf v}]$ can be related to the orbital momentum of the proton with respect to the centre of Earth,
\begin{equation}
{\boldsymbol L}_{\oplus}\equiv \gamma m \big[{\bf R}_{\oplus}\times {\bf v}\big]=\big[{\bf R}_{\oplus}\times {\boldsymbol p}\big]\, . \label{OrbitalMomentum}
\end{equation}
Then, this GR correction superficially resembles the  Lense-Thirring interaction
\cite{Lense,Schiff} between the orbital  momentum of the proton
with respect to the Earth, ${\boldsymbol L}_{\oplus}$, and the spin of the central body,
${\boldsymbol S}_{\oplus}\equiv IM_{\oplus}R^2_{\oplus}{\boldsymbol\omega}$:
\begin{equation}
\frac{3\gamma\big({\bf v}[{\boldsymbol\omega}
	,\,{\bf r}]\big)}{4c^2}=
\frac{3}{4mc^2}\cdot\frac{({\boldsymbol S}_{\oplus}{\boldsymbol L}_{\oplus})}{IM_{\oplus}R^2_{\oplus}}\,.
\label{SpinCB}
\end{equation}
For a particle in a cyclic accelerator
${\boldsymbol L}_{\oplus}$ is not conserved, though. The last term in the curly brackets in Eq. (\ref{Cycl30})
is the GR correction   $\propto {\boldsymbol L}_{\oplus}$.

\subsection{Relativistic Dynamics of Spinning Particles with  MDM and EDM}


Let  ${\boldsymbol\mu}$ and ${\bf d}$  be the MDM and EDM, respectively,
of the particle in the "falling lift" frame $K_0$ (see (\ref{G160}))
with the Riemann normal coordinates. In this case, the equation for the precession of the polarization vector
${\boldsymbol S}$  has the familiar nonrelativistic form
\begin{equation}
\frac{\d{\boldsymbol S}}{\d t}=\frac{2\mu}{\hbar}\big[{\boldsymbol S}\times {\bf H}\big]+\frac{2d}{\hbar}\big[{\boldsymbol S}\times {\bf E}\big].
\label{SD10}
\end{equation}

In the standard description of the spin dynamics of the relativistic particle in the laboratory reference
frame, one introduces the polarization 4-vector $P^a$ such that $P^a=(0,\,{\boldsymbol S})$
in the reference frame  $K_0$. For a particle moving with the 4-velocity
${u^a}$, the Lorentz transformation gives
\begin{equation}
  {\boldsymbol P}={\boldsymbol S}+\frac{\gamma^2}{c^2(\gamma+1)}({\boldsymbol S}{\bf v}){\bf v}, \quad
P^0=\frac{\gamma}{c}({\boldsymbol S}{\bf v}),
\quad u_a P^a=0.
\label{SD20}
\end{equation}
The corresponding relativistic equation of motion for spin  has the form \cite{nelson1959search,fukuyama2013derivation}
\begin{gather}
\frac{\d P^a}{\d s}+(\gamma^a_{bc}u^c)P^b=\frac{2}{\hbar c}\Big\{\mu F^a_{\  b}P^b-\mu' u^aF^{b}_{\  c}u_bP^c
-d\cdot \tilde{F}^{a}_{\  b}P^b+d\cdot u^a\tilde{F}^{b}_{\  c}u_bP^c\Big\},
\label{SD30}
\end{gather}
where
\begin{gather}
\tilde{F}^{ab}\equiv\frac12\varepsilon^{abcd}F_{cd}, \quad \varepsilon^{0123}=1,
\nonumber \\
 \mu\equiv(G+1)\cdot\frac{q\hbar}{2mc}, \quad \mu'=G\frac{q\hbar}{2mc}, \quad G=(g-2)/2,
\nonumber \\
d=\eta_{\rm EDM}\frac{q\hbar}{2mc}.
\label{SD40}
\end{gather}

\subsection{Frenkel-Thomas-Bargmann-Michel-Telegdi equation in presence of gravity}

Relativistic equation (\ref{SD30})  is not yet a final result; we need  precession of the experimentally
measured polarization vector ${\boldsymbol S}$ defined in the comoving frame.  We consider contributions to angular velocity of spin precession from the MDM, EDM and
GR, correspondingly.

\subsubsection{Spin rotation due the MDM}

This contribution was obtained first in classical works by Frenkel \cite{Frenkel}  and Thomas \cite{Thomas}
. The standard representation used in accelerator physics is due to Bargmann, Michel and Telegdi  \cite{bargmann1959precession}:
\begin{gather}
{\boldsymbol\Omega}_{MDM}=-\frac{2\mu+2\mu'(\gamma-1)}{\gamma\hbar}{\bf H}+\frac{2\mu'\gamma}{(\gamma+1)\hbar}({\bf H}{\boldsymbol\beta}){\boldsymbol\beta}
+\frac{2\mu+2\mu'\gamma}{(\gamma+1)\hbar}[{\boldsymbol\beta}\times {\bf E}]
\nonumber \\
=-\frac{q}{mc}\left\{\left(G+\frac{1}{\gamma}\right){\bf H}-
\frac{\gamma G}{\gamma+1} ({\bf H}{\boldsymbol\beta}){\boldsymbol\beta}
-\left(G+\frac{1}{\gamma+1}\right)[{\boldsymbol\beta}\times {\bf E}]\right\}.
\label{SD60}
\end{gather}

\subsubsection{Spin rotation due the EDM}

The first discussion of the search for EDM  spin precession from the EDM was previously discussed in
\cite{nelson1959search,fukuyama2013derivation}.  According to Eq. (\ref{SD30}), this contribution  readily derives from Eq. (\ref{SD60}) for upon substitutions $\mu\longrightarrow-d$, $\mu'\longrightarrow-d$,
${\bf E}\longrightarrow {\bf H}$, and ${\bf H}\longrightarrow -{\bf E}$:
\begin{equation}
{\boldsymbol\Omega}_{\rm EDM}=-\eta_{\rm EDM}\frac{q}{mc}\left\{{\bf E}-\frac{\gamma}{\gamma+1}({\bf E}{\boldsymbol\beta}){\boldsymbol\beta}+
[{\boldsymbol\beta}\times {\bf H}] \right\}.
\label{SD130}
\end{equation}

\subsubsection{Gravity driven spin rotation}


Make in Eq. (\ref{SD60}) substitutions
$2\mu =q\hbar/mc$,  $\mu'=d=0$ and \cite{pomeranskiiJETP}
\begin{equation}
\frac{q}{mc^2}{\bf E}^{\alpha}\longrightarrow (\gamma_{\alpha0c}u^c), \quad
\frac{q}{mc^2}{\bf H}^{\alpha}\longrightarrow \frac12\varepsilon_{\alpha\beta\rho}(\gamma_{\beta\rho c}u^c)
\label{SD140}
\end{equation}
with the result
\begin{gather}
{\boldsymbol\Omega}^{GR}
=\frac{2\gamma+1}{\gamma+1}\frac{[{\boldsymbol\beta}\times {\boldsymbol g}_0]}{c}
\nonumber \\
-\Bigg(\frac{2\gamma-1}{\gamma}\left(1-\frac{r_g}{2R_{\oplus}}\right)+\frac{5\gamma+3}{2(\gamma+1)c}\big({\boldsymbol\omega}[{\boldsymbol R}\times {\boldsymbol \beta}]\big)\Bigg){\boldsymbol\omega}
\nonumber \\
+\frac{\gamma}{\gamma+1}\Bigg(({\boldsymbol
\omega}{\boldsymbol\beta})-\frac{r_g}{R_{\oplus}}\left(\frac12({\boldsymbol\omega}{\boldsymbol\beta})-
({\boldsymbol\omega}{\bf n}_{\oplus})({\boldsymbol\beta}{\bf n}_{\oplus})\right)\Bigg){\boldsymbol\beta}
+\frac{\gamma}{(\gamma+1)c}
({\boldsymbol\omega}{\boldsymbol\beta})[{\boldsymbol\omega}\times {\bf R}]
\nonumber \\
-\frac{2\gamma-1}{\gamma}\cdot\frac{r_g}{R_{\oplus}}({\boldsymbol\omega}{\bf n}_{\oplus}){\bf n}_{\oplus}
\\
-\frac{r_gI}{2R_{\oplus}}\bigg(\frac{2\gamma-1}{\gamma}\Big({\boldsymbol\omega}-
3({\boldsymbol\omega}{\bf n}_{\oplus}){\bf n}_{\oplus}\Big)-
\frac{\gamma}{\gamma+1}\Big(({\boldsymbol\omega}{\boldsymbol\beta})-3({\boldsymbol\omega}{\bf n}_{\oplus})({\boldsymbol\beta}{\bf n}_{\oplus})\Big){\boldsymbol\beta}\bigg). \nonumber
\label{SD150}
\end{gather}
Here the first term is de Sitter geodetic effect \cite{deSitter,pomeranskiiJETP,pomeranskiiUFN}. The term $\propto {\boldsymbol\omega}$ is the Earth rotation effect. Note the relativistic factor of
$(2\gamma -1)/\gamma$ and GR corrections, including the one $\propto \big({\boldsymbol\omega}[{\boldsymbol R}
\times {\boldsymbol \beta}]\big) $, already encountered in the cyclotron rotation, Eq. (\ref{Cycl30}). The third term, $\propto {\boldsymbol\beta}$, describes spin precession about the momentum, including the precession about the velocity $[{\boldsymbol\omega}\times {\bf R}]$  from Earth's rotation. All terms $\propto ({r_gI}/{R_{\oplus}}){\boldsymbol\omega}$ derive from components of the Kerr metric proportional to ${\boldsymbol\omega}$ (see Appendix B). The last line is the  relativistic generalization of the Lense-Thirring (LT) contribution to spin precession \cite{Lense}. Indeed, in terms of the spin of the Earth,
${\boldsymbol S}_{\oplus}\equiv IM_{\oplus}R^2_{\oplus}{\boldsymbol\omega}$, the  last line of Eq. (\ref{SD150})  takes the form
\begin{equation}
\left(\delta{\boldsymbol\Omega}^{GR}_{LT}\right)^{\alpha}=-\frac{k}{R_{\oplus}^3}
\left(\frac{2\gamma-1}{\gamma}\delta^{\alpha\gamma}-\frac{\gamma}{\gamma+1}{\boldsymbol\beta}^{\alpha}
{\boldsymbol\beta}^{\gamma}\right)
\Big({\boldsymbol S}_{\oplus}^{\gamma}-3({\boldsymbol S}_{\oplus}{\bf n}_{\oplus}){\bf n}_{\oplus}^{\gamma}\Big).
\label{SD151}
\end{equation}

\subsubsection{The subtracted angular velocity of spin precession }

In storage ring experiments, one is interested in  the so-called subtracted angular velocity of spin precession ${\boldsymbol{\Omega}}_{s}$,
\begin{gather}
{\boldsymbol{\Omega}}_{s}\equiv{\boldsymbol\Omega}_{MDM}+{\boldsymbol\Omega}_{\rm EDM}
+{\boldsymbol\Omega}^{GR}-{\boldsymbol\Omega}_c
\nonumber \\
=-\frac{q}{mc}\left\{G{\bf H}-\frac{\gamma}{\gamma+1}\left(G-\frac{1}{\gamma-1}\right)({\bf H}{\boldsymbol\beta}){\boldsymbol\beta}-
\left(G-\frac{1}{\gamma^2-1}\right)[{\boldsymbol\beta}\times {\bf E}]\right\}
\nonumber \\
-\eta_{EDM}\frac{q}{2mc}\left\{{\bf E}-\frac{\gamma}{\gamma+1}({\bf E}{\boldsymbol\beta}){\boldsymbol\beta}+
[{\boldsymbol\beta}\times {\bf H}] \right\}
\nonumber \\
+{\boldsymbol{\Omega}}_s^{GR},
\label{SD160}
\end{gather}
This subtracted  ${\boldsymbol\Omega}_s$ describes the spin precession as seen by the comoving observer. Of prime interest is the GR term \cite{VergelesJETP}:
\begin{gather}
{\boldsymbol{\Omega}}^{GR}_{s}=-
\frac{1}{\gamma{\boldsymbol\beta}^2c}[{\boldsymbol\beta}\times {\boldsymbol g}_0]+
\frac{1}{\gamma}\left(1-\frac{r_g}{2R_{\oplus}}(1-I)\right){\boldsymbol\omega}+
\frac{1}{\gamma{\boldsymbol\beta}^2c}
\big({\boldsymbol\omega}[{\bf R}\times{\boldsymbol\beta}]\big){\boldsymbol\omega}
\nonumber \\
-\frac{1}{\gamma{\boldsymbol\beta}^2 c}({\boldsymbol\omega}{\boldsymbol\beta})
[{\boldsymbol\omega}\times {\bf R}]
+\frac{1}{\gamma}\cdot\frac{r_g}{R_{\oplus}}\left(1-\frac32I\right)({\boldsymbol\omega}{\bf n}_{\oplus}){\bf n}_{\oplus}-
\nonumber \\
-\frac{\gamma+1}{\gamma{\boldsymbol\beta}^2}
\left\{ \left(1-\frac{r_g}{2R_{\oplus}}(1-I) +\frac{3\gamma}{2c(\gamma+1)}
\big({\boldsymbol\omega}[{\bf R}\times{\boldsymbol\beta}]\big)\right)
({\boldsymbol\omega}{\boldsymbol\beta})
\right.
\nonumber \\
\left.
+\frac{r_g}{R_{\oplus}}\left(1-\frac32I\right)({\boldsymbol\omega}{\bf n}_{\oplus})({\boldsymbol\beta}{\bf n}_{\oplus})\right\}
{\boldsymbol\beta}.
\label{SD170}
\end{gather}
In the storage ring experiments, it is not the full story yet.

\subsection{Compensation of the gravity pull and gravity as a Standard Candle}

Silenko and Teryaev were the first to realize in 2007 that a condition of the closed orbit in storage rings calls for compensation of the Earth's gravity pull by the focusing EM fields \cite{Silenko}. They condidered all magnetic storage rings with MDM rotations dominated by magnetic fields, when the impact of the focusing fields on the spin precession is negligibly small forl all the practical purposes. Much more interesting is the case of all electric, frozen-spin storage rings  treated in 2012 by Orlov, Flanagan and Semertzidis \cite{Orlov} and in 2016 by Obukhov, Silenko and Teryaev \cite{Obukhov}.

Decompose the electric field into the
confining pure radial ${\bf E}_0$,  and the focusing ${\bf E}_f$, which  compensates the GR contribution into the orbital motion:
\begin{equation}
\frac{\d u^{\alpha}}{\d t}=
\frac{q}{mc}{\bf E}_0^{\alpha}+\bigg\{c\left[(\gamma_{\alpha 0c}u^c)+(\gamma_{\alpha\beta c}u^c){\boldsymbol\beta}^{\beta}\right]+\frac{q}{mc}{\bf E}_f^{\alpha}\bigg\}.
\label{RC10}
\end{equation}
The compensation means vanishing term in the curly braces. Upon some exercise with connections, one finds
\begin{gather}
\frac{q}{mc}{\bf E}_{f}=\gamma\Bigg\{-\frac{(2\gamma^2-1){\boldsymbol g}_0}{\gamma^2 c}
+\frac{1}{c}\big[{\boldsymbol\omega}\times [{\boldsymbol\omega}\times {\bf R}]\big]
\nonumber \\
+\left(2-\frac{r_g}{R_{\oplus}}(1-I)+\frac{3}{2c}\big(
{\boldsymbol\beta}[{\boldsymbol\omega}\times {\bf R}]\big)\right)[{\boldsymbol\omega}\times {\boldsymbol\beta}]-
\frac{r_g}{R_{\oplus}}(2-3I)({\boldsymbol\omega}{\bf n}_{\oplus})[{\boldsymbol\beta}\times {\bf n}_{\oplus}]\Bigg\}.
\label{RC20}
\end{gather}
It is crucial  that in view of the
$({\boldsymbol\beta}{\boldsymbol g}_0)=(\boldsymbol{\beta}{\bf n}_\oplus)=0$, the loop integral along the storage ring $\oint{\bf E}_f d{\bf r}=0$, {\it i.e.,} the field  ${\bf E}_f$  is potential one and can be implemented electrostatically \cite{VergelesJETP}

Therefore, the net GR contribution to spin precession boils down to rotation of the MDM in the motional magnetic field from the focusing electric field ${\bf E}_{f}$
given by Eq. (\ref{SD60}) plus the gravity proper term (\ref{SD150}):
\begin{gather}
{\boldsymbol\Omega}_f^{GR}= \left(G+\frac{1}{\gamma+1}\right)\frac{q}{mc}[\boldsymbol{\beta}\times\boldsymbol{E}_f] +\boldsymbol{\Omega}^{Gr}_s = 	{\boldsymbol\Omega}_g^{GR}	+{\boldsymbol\Omega}_{\boldsymbol\omega}^{GR}
\nonumber \\
= \frac{1-(2\gamma^2-1)G}{\gamma}\frac{|{\bf g}_0|}{c} [ {\bf n}_{\oplus}\times {\boldsymbol\beta}]	
\nonumber \\
-\gamma \frac{G\gamma}{c}({\boldsymbol\omega}{\boldsymbol\beta})\cdot
[{\boldsymbol\omega}_t\times{\bf R}]	
\nonumber \\
+\Bigg[-\frac{1-2(\gamma^2-1)G}{\gamma}
\left(1-\frac{r_g}{2R_{\oplus}}(1-I)\right)
\nonumber \\
+\frac{1}{2\gamma}\bigg(\left(5\gamma^2-3\right)G-3\bigg)\frac{1}{c}
\big({\boldsymbol\omega}_t [{\bf R}\times{\boldsymbol\beta}]\big)\Bigg]{\boldsymbol\omega}
\nonumber \\
+\left(\frac{2(\gamma^2-1)}{\gamma}G-\frac{1}{\gamma}\right)\frac{r_g}{R_{\oplus}}
\left(1-\frac{3}{2} I \right)({\boldsymbol\omega}{\bf n}_{\oplus}){\bf n}_{\oplus}
\nonumber \\
-\Bigg[\left(2\gamma G+\frac{\gamma}{\gamma+1}\right)\left(1-\frac{r_g}{2R_{\oplus}}(1-I)\right)
\nonumber \\
+\left(\gamma G+\frac{\gamma}{\gamma+1}\right)
\frac{3}{2c}\big({\boldsymbol\omega}_t [{\bf R}\times{\boldsymbol\beta}]\big)\Bigg]
({\boldsymbol\omega}_t{\boldsymbol\beta}){\boldsymbol\beta}.
\label{RC40}
\end{gather}
Here we neglected the correction from ${\bf E}_{f}$ to the EDM effect.

The same result can be obtained from general formula (\ref{SD160}) by setting ${\bf H}=0$ and lumping together contributions from ${\boldsymbol{\Omega}}^{GR}_s$ and ${\bf E}_{f}$ according to Eqs. (\ref{SD170})
and (\ref{RC20}),  respectively.

A beauty of the all electric frozen spin ring in flat space is that the
spin
would precess around the radial electric field entirely
due to the EDM:
\begin{equation}
{\boldsymbol\Omega}_{EDM}=-\eta_{EDM}\frac{q}{mc}{\bf E}_0 = \eta_{EDM}\frac{q}{mc}{E}_0 {\bf e}_r\, .
\label{ClOrb20}
\end{equation}
It is obvious that gravity pull term  in (\ref{RC40}),
\begin{equation}
{\boldsymbol\Omega}_g^{GR}= \frac{1-G(2\gamma^2-1)}{\gamma} \cdot \frac{|{\bf g}_0|}{c}[\boldsymbol{n}_{\oplus}\times \boldsymbol{\beta}] =\frac{1-G(2\gamma^2-1)}{\gamma} \cdot \frac{|{\bf g}_0|}{c}\beta {\bf e}_r \, ,
\label{eq:fakeEDM}
\end{equation}
generates the fake EDM signal. This formula for arbitrary energies was obtained in 2016 by Obukhov, Silenko, and Teryaev
\cite{Obukhov}. In the all electric frozen spin ring
\begin{equation}
\beta^2=\frac{1}{1+G}
\label{ClOrb10}
\end{equation}
and ({eq:fakeEDM}) coincides
\cite{nikolaev2019gravity} with  the  2012 result by Orlov, Flanagan, and Semertzidis
\cite{Orlov}:
\begin{equation}
{\boldsymbol\Omega}^{GR}_g= \sqrt{G} \frac{|{\bf g}|_0}{c}\boldsymbol{e}_{r}.
\label{RC50}
\end{equation}
A sensitivity to true proton EDM depends on the confining electric field $E_0$. Designs of all electric frozen-spin rings aim at electric fields up to challenging  $E_0 \equiv 10$ Mev/m. In such a ring, the gravity pull effect amounts to $d_p^{fake} \equiv 28.8\cdot 10^{-29}$ e cm, {\it i.e.,} $\eta_{EDM} \equiv 27.4 \cdot 10^{-15}$. At weaker electric field of $E_0 \equiv 2.5$ Mev/m the sensitivity to the EDM will be lower and the gravity pull effect would amounts to $d_p^{fake} \equiv 115\cdot 10^{-29}$ e cm, {\it i.e.,} $\eta_{EDM} \equiv 110 \cdot 10^{-15}$ (see \cite{abusaif2019storage}, Appendix D).

Fortunately enough, all electric rings are ideally suited to run with counter-rotating beams. While the EDM effect will be identical for the anticlockwise (ACW) and clockwise (CW) beams and can easily be separated from the gravity pull effect which is of the opposite sign for the CW and ACW beams. Furthermore, the gravity pull effect can be taken for the Standard Candle to calibrate the ring performance (see \cite{abusaif2019storage}, Appendix D).

We defer a scrutiny of ${\boldsymbol\Omega}^{GR}_{\boldsymbol\omega}$ and of importance of the vertical component of ${\boldsymbol \omega}$  to Section 4.4.

\section{Maxwell equations for  pure electrostatic system in non-inertial frames }



So far we focused on spin dynamics in a given external electromagnetic field. However, in curved and non-inertial space-time, the electromagnetic field itself  must be determined using Maxwell's equations for the specified charge distributions and currents. The salient feature of non-inertial space-times with the stationary metric is a nonvanishig off-diagonal elements,
\begin{equation}
g_{0i}\neq 0.
\label{intr10}
\end{equation}
As as we shall see, these give rise to nonvanishing geometric magnetic field in the pure electrostatic systems residing at rest on rotating bodies \cite{VergelesJETP,VergelesJHEP}. Weak though the GR effects are, the GMF has quite an impact in ultrahigh precision spin dynamics of the EDM experiments.

By the definition of the electrostatic system at rest in the laboratory frame, the 3-current vanishes:
\begin{equation}
(\partial/\partial x^0)J^{\mu}(x)=0, \quad J^0\neq 0, \quad J^i=0\, .
\label{intr20}
\end{equation}
The subsequent derivation of the GMF  proceeds in this reference frame and this very special
4-current $J^{\mu}\ $  \cite{VergelesJETP,VergelesJHEP}.

Any antisymmetric field in the three-dimensional space $\sqrt{-g}F^{ij}=-\sqrt{-g}F^{ji}$ can be represented as
\begin{equation}
\sqrt{-g}F^{ij}=\varepsilon_{ijk}\partial_k\psi+\left(\partial_i{\cal A}_j-\partial_j{\cal A}_i\right)
\label{H10n}
\end{equation}
with pseudoscalar potential $\psi$. Technically, in the rotating frames one faces the formal issue of the horizon.
 However, in all the cases of practical interest, the charge and current distributions are localized well inside the horizon radius. Consequently, for fields decreasing at infinity, in three-dimensional Euclidean space the decomposition (\ref{H10n}) is well defined and unique, and $J_i=0$ entails ${\cal A}_i=0$.

According to the rules (\ref{G64}), the field definition (\ref{M50}) and  the fact that $\tilde{e}^i_0=0$, we obtain:
\begin{equation}
F^{ij}=\tilde{e}^i_a\tilde{e}^j_bF^{ab}=-\varepsilon_{\alpha\beta\gamma}
\tilde{e}^i_{\alpha}\tilde{e}^j_{\beta}{\bf H}^{\gamma}=-\frac{e^0_0}{\sqrt{-g}}
\varepsilon_{ijk}e^{\alpha}_k{\bf H}^{\alpha}.
\label{H20n}
\end{equation}
Here we  made use of the relations (\ref{H25n}) in Appendix C.
Then Eq. (\ref{H10n}) with ${\cal A}_i=0$ in conjunction with Eq. (\ref{H20n}) yields
\begin{equation}
{\bf H}^{\alpha}=-(e^0_0)^{-1}\tilde{e}^i_{\alpha}\partial_i\psi.
\label{H30n}
\end{equation}

Now we turn to the  homogeneous Maxwell equations (\ref{M20}).
We express the  holonomic field $F_{\mu \nu}$  in terms of the electric  and  magnetic fields:
\begin{equation}
F_{i0}=e^a_ie^b_0F_{ab}=e^0_0e^{\alpha}_iF_{\alpha0}=
-e^0_0e^{\alpha}_i{\bf E}^{\alpha}=\partial_iA_0, \quad
\mbox{or}
\quad {\bf E}^{\alpha}=-\frac{\tilde{e}^i_{\alpha}}{e^0_0}F_{i0},
\label{H40n}
\end{equation}
\begin{equation}
F_{ij}=e^a_ie^b_jF_{ab}=-\varepsilon_{\alpha\beta\gamma}e^{\alpha}_ie^{\beta}_j{\bf H}^{\gamma}-
\left(e^0_je^{\alpha}_i-e^0_ie^{\alpha}_j\right){\bf E}^{\alpha}.
\label{H50n}
\end{equation}
Eq. (\ref{M20}) with $\mu=0$ implies the identity $\varepsilon_{ijk}\partial_kF_{ij}=0$. Therefore,
applying the operator $\varepsilon_{ijk}\partial_k$ to the Eq. (\ref{H50n}), using (\ref{H30n})
and the identity $\varepsilon_{ijk}\partial_kF_{i0}=-\varepsilon_{ijk}\partial_k\left(
e^0_0e^{\alpha}_i{\bf E}^{\alpha}\right)=0$ (see (\ref{H40n})), we
obtain the equation
\begin{equation}
\partial_i\bigg(\sqrt{-g}\left(e^0_0\right)^{-2}g^{ij}\partial_j\psi\bigg)=
-\varepsilon_{ijk}e^0_0e^{\alpha}_i{\bf E}^{\alpha}\partial_k\left(\frac{e^0_j}{e^0_0}\right).
\label{H60n}
\end{equation}
Recall that  $e^0_j=g_{0j}/\sqrt{g_{00}}\neq0$  according to (\ref{tetrad}) and (\ref{intr10}). Therefore
if ${\bf E}\neq0$, then (\ref{H60n}) entails  $\partial_j\psi\neq0$ and,
according to Eq. (\ref{H30n}),  ${\bf H}\neq0$ as well. We shall refer to the field (\ref{H30n}) as the geometric magnetic field $ {\bf H}_{\boldsymbol\omega}$, the subscript ${\boldsymbol\omega}$ is a reference to its origin in the rotation of the laboratory frame $K$.

Next we look into the equation for the electric field.
Let's  express $F^{i0}$ in terms of physical fields:
\begin{equation}
F^{i0}=\tilde{e}^i_a\tilde{e}^0_bF^{ab}=\tilde{e}^0_0\tilde{e}^i_{\alpha}{\bf E}^{\alpha}-
\varepsilon_{\alpha\beta\gamma}\tilde{e}^i_{\alpha}\tilde{e}^0_{\beta}{\bf H}^{\gamma}
=\frac{1}{e^0_0}\tilde{e}^i_{\alpha}{\bf E}^{\alpha}-\frac{1}{\sqrt{-g}e^0_0}\varepsilon_{ijk}e^0_j\partial_k\psi.
\label{H70n}
\end{equation}
The substitution of the right-hand side of (\ref{H70n}) into Eq. (\ref{M30}) with $\mu=0$ leads to
\begin{equation}
\partial_i\left(\frac{\sqrt{-g}}{e^0_0}\tilde{e}^i_{\alpha}{\bf E}^{\alpha}\right)-
\varepsilon_{ijk}\partial_i\left(\frac{e^0_j}{e^0_0}\right)\cdot\partial_k\psi=
4\pi\sqrt{-g}J^0.
\label{H80n}
\end{equation}
Here we have used representation (\ref{H30n}) and one of the relations (\ref{H25n}).
The system of equations  (\ref{H30n}), (\ref{H60n}) and (\ref{H80n})  is complete  and exact. The electric field ${\bf E}$  is defined in Eq.  (\ref{H40n})  in terms of the potential $A_0$.

The curved space-time of the rotating body is described  by the Kerr  metric (see Appendix B).
Making use of (\ref{metric20}) and (\ref{IM}), one can readily derive a hierarchy of the expansion of electromagnetic files in small parameters
$r_g$, $r_g{\boldsymbol{\omega}}$, ${\boldsymbol{\omega}}\otimes{\boldsymbol{\omega}}$.
%
To the zeroth order,  
the potential $A_0^{(0)}$ and the corresponding Minkowski space defined electric field  are related as ${\boldsymbol{E}}^{(0)}=-\nabla A_0^{(0)}$. Concerning the geometric magnetic field,  the off-diagonal $g_{0i}$ of Eq. (\ref{metric20}),  $e^0_i$  of Eq. (\ref{tetrad}) and  $\tilde{e}^0_{\alpha}$ of Eq. (\ref{ONB})  are all $\propto {\bf \omega} $.  Consequently  $e^0_j/e^0_0=\O({\boldsymbol\omega})$ and, according to (\ref{H60n}), the expansion for the magnetic potential $\psi$ starts with the linear term $\psi^{(1)}=\O({\boldsymbol{\omega}})$.

According to the expansion of the Kerr metric in Appendix B, we have
\begin{equation}
\frac{\sqrt{-g}\tilde{e}^i_{\alpha}}{e^0_0}=\left(1+\frac{r_g}{R}\right)\delta^i_{\alpha}
+\left(\frac{[{\boldsymbol\omega}\times{\bf R}]^2}{2c^2}\delta^i_{\alpha}-
\frac{[{\boldsymbol\omega}\times {\bf R}]^i[{\boldsymbol\omega}\times{\bf R}]^{\alpha}}{2c^2}\right)\, ,
\label{H90n}
\end{equation}
which does not contain the linear term.  Then, by virtue of equation (\ref{H80n}), the electric field  acquires the first correction only to the second order in ${\boldsymbol{\omega}}$, i.e., $ {\bf E}^{(1)}=0, \,{\bf E}^{(2)}\neq 0$. In the due turn,  Eq. (\ref{H60n})  guarantees that the quadratic correction to the magnetic potential vanishes: $\psi^{(2)}=0$.

For a tedious derivation of ${\bf E}^{(2)}$ we refer the readers to Ref. \cite{VergelesJETP} and concentrate instead on ${\bf H_\omega}$. The rotating body of practical interest is the Earth. It is the case of weak gravity. On the terrestrial surface
$r_g/R_{\oplus}\sim 10^{-9}$  and $\omega R_{\oplus}/c \sim 1.5\cdot 10^{-6}$. Hence we keep the terms $\O(|{\boldsymbol\omega}|)$.
To this approximation,
equations  (\ref{metric20})-(\ref{Connection}) simplify to
\begin{gather}
g_{00}=g^{00}=1, \quad g_{0i}=g^{0i}=-\frac{\big[{\boldsymbol\omega}\times{\bf R}]^i}{c},
\nonumber \\
g_{ij}=g^{ij}=-\delta^{ij}, \quad g=-1,
\nonumber \\
e^0_0=1, \quad e^0_i=-\frac{\big[{\boldsymbol\omega}\times{\bf R}\big]^i}{c},
\quad e^{\alpha}_i=\delta^{\alpha}_i, \quad e^{\alpha}_0=0,
\nonumber \\
\tilde{e}^0_0=1, \quad \tilde{e}^0_{\alpha}=\frac{\big[{\boldsymbol\omega}\times {\bf R}\big]^{\alpha}}{c},
\quad \tilde{e}_{\alpha}^i=\delta_{\alpha}^i, \quad \tilde{e}^i_0=0,
\nonumber \\
\frac{c}{2}\varepsilon_{\alpha\beta\rho}\gamma_{\beta\rho 0}={\boldsymbol\omega}^{\alpha}.
\label{H90}
\end{gather}
The second order corrections to the electric potential and electric field can be neglected and we have the familiar ${\bf E}=-\nabla A_0$ and  the Poisson equation $\div{\bf E}=-\Delta A_0=4\pi J^0$, while the Poisson equation for the potential of the geometric magnetic field takes a simple form,
\begin{equation}
\Delta\psi=\frac{2}{c}({\boldsymbol\omega}{\bf E})\,,
\label{H140}
\end{equation}
to be used in the subsequent analysis of terrestrial experiments. Note manifest conservation of parity with pseudoscalar $\psi$.

\section{Experimental manifestations of the geometric magnetic field}


\subsection{Geometric magnetic field of the spherical shell}

The phenomenon of geometric magnetic field is best illustrated by the charged spherical shell at rest in the  laboratory frame, when
\begin{equation}
{\bf E}({\bf r})=\left\{
\begin{array}{rl}
\dfrac{Q{\bf r}}{r^3},  &  \mbox{for} \quad r>a, \\  [4mm]
0, & \mbox{for} \quad r<a.
\end{array}  \right.
\label{IE10}
\end{equation}
Here $Q$ is the charge of the shell, and the  radius-vector ${\bf r}=0$ at the centre of the  shell.

A straightforward solution to Eq. (\ref{H140}) is
\begin{equation}
\psi({\bf r})=-\frac{\boldsymbol\omega}{2\pi c}\int\d^{(3)}x\frac{1}{|{\bf r}-{\bf x}|}{\bf E}({\bf x})=
-\frac{Q\cdot ({\boldsymbol\omega}{\bf r})}{3c}\left\{
\begin{array}{rl}
\dfrac{3}{r}-\dfrac{a^2}{r^3},  &  \mbox{for} \quad r>a, \\  [4mm]
\dfrac{2}{a}, & \mbox{for} \quad r<a\,.
\end{array}  \right.
\label{IE30}
\end{equation}
The corresponding CMF admits a crystal clear interpretation,
\begin{equation}
{\bf H}_{\boldsymbol\omega}({\bf r})=\left\{
\begin{array}{rl}
\dfrac{1}{r^3}\Big\{3({\boldsymbol\mu}{\bf n}){\bf n}-{\boldsymbol\mu}\Big\}+
\dfrac{1}{c}\big[{\bf E}({\bf r})\times [{\boldsymbol\omega}\times {\bf r}]\big],  &  \mbox{for} \quad r>a, \\ [4mm]
\dfrac{2Q{\boldsymbol\omega}}{3ca}, & \mbox{for} \quad r<a,
\end{array}  \right.
\label{IE40}
\end{equation}
where ${\bf n}={\bf r}/{r}$ and
\begin{equation}
{\boldsymbol\mu}=\frac{Qa^2}{3c}{\boldsymbol\omega}
\nonumber
\end{equation}
is the geometric magnetic moment  of  the charged shell, induced by the Earth's rotation. The second term in (\ref{IE40}) is the
familiar motional magnetic field in the rotating laboratory frame $K$ which is entailed by the electric field in the inertial frame $K'$.

\subsection{False EDM signal in the neutron EDM experiments }

In the neutron EDM experiments, ultracold neutrons are subjected to uniform electric and magnetic fields. The principal observable is the change of the Larmor precession frequency
\begin{equation}
f_n =\frac{1}{\pi\hbar}|\mu_n{\bf B} +d_n{\bf E}| \label{eq:Larmor}
\end{equation}
from the parallel to anti-parallel magnetic and electric fields. The EDM is extracted from the frequency shift
\begin{equation}
d_n = \frac{\pi \hbar \Delta f}{2|{\bf E}|} .
\label{eq:EDM extraction}
\end{equation}
The implicit assumption is that flipping the electric field does not change the magnetic one, which is not the case with the geometric magnetic field.

In practice the electric field is generated in the plane capacitor with the gap much narrower than the size of two plane electrodes.
In the gap in between the plates one has
\begin{equation}
{\bf E}_0=(0,\,0,\,{\cal E}_0)=-\nabla A_0(z), \quad A_0(z)=-{\cal E}_0 z,
\label{IIiE10}
\end{equation}
while beyond the gap the electric field vanishes.
Now we solve the Poisson equation (\ref{H140}) for the magnetic potential, representing the electric field through $A_0(z)$ and integrating by parts:
\begin{equation}
\psi(z)=\frac{2{\boldsymbol\omega}_z}{c}\int\d z'\Delta^{-1}(z-z'){\cal E}_0
=-\frac{2{\boldsymbol\omega}_z}{c}\int\d z'\frac{\d}{\d z}\Delta^{-1}(z-z')A_0(z'),
\label{IIiE20}
\end{equation}
where $\Delta^{-1}(z)=|z|/2$ is the inverse to the Laplace operator. One more differentiation yields \begin{equation}
{\bf H}_{\boldsymbol\omega}=-\nabla\psi(z)
=\left(0,\,0,\,\frac{2{\boldsymbol\omega}_z}{c}A_0(z)\right)=
\left(0,\,0,\,-\frac{2{\boldsymbol\omega}_z{\cal E}_0}{c}z\right)=
-\frac{2{\boldsymbol\omega}_zz}{c}{\bf E}_0.
\label{IIiE30}
\end{equation}
The geometric  field ${\bf H}_{\boldsymbol\omega}$ is parallel to the external electric field ${\bf E}_0$. Its salient feature is the nonvanishing constant gradient
\begin{equation}
\frac{d{\bf H}_{\boldsymbol\omega}}{d z} = -\frac{2{\boldsymbol\omega}_z}{c} {\bf E}_0 \, .
\end{equation}

The crucial component of the neutron EDM experiments is the comagnetometry: one measures the neutron spin precession frequency with respect to that of the mercury comagnetometer. The mercury atoms are uniformly distributed in the volume of the neutron storage cell, and the average geometric magnetic field acting on the mercury comagnetometer vanishes: $\langle {\bf H}_{\boldsymbol\omega}^{(Hg)}\rangle =
{\bf H}_{\boldsymbol\omega}(0) =0$.
The centre of mass of neutrons differs from that of the mercury by the  offset  $\langle z \rangle$, what entails the nonvanishing average geometric magnetic field acting on the magnetic moment of neutrons
\begin{equation}
{\bf H}_{\boldsymbol\omega}^{(n)} = -\frac{2\langle z\rangle{\boldsymbol\omega}_z}{c}{\bf E}_0 .
\label{eq:HgeomNeutron}
\end{equation}

The most important point is that this geometric field changes the sign when the electric field is flipped. The net effect is that the apparent EDM of neutrons, $d_n^{obs}$, as given by the procedure (\ref{eq:EDM extraction}), will acquire the false component, $d_{n}^{obs} = d_n + d_{false}$, where
\begin{equation}
d_{false} = -\frac{2\langle z \rangle{\boldsymbol\omega}_z}  {c}\mu_n\, . \label{eq:dFalse}
\end{equation}
In the experiment \cite{pendlebury2015revised} the neutron centre of mass offset was $\langle z \rangle \simeq 2.8$mm, the more recent experiment \cite{abel2020measurement} reports $\langle z \rangle \simeq 3.9$mm . Taking the former, we find $d_{false} \approx 2.5\times  10^{-28}$ e$\cdot$cm. It is still way below the recently reported \cite{abel2020measurement}  best result for the neutron EDM, $d_n = (0.0\pm 1.1_{stat} \pm 0.2_{sys})\times 10^{-26}$e$\cdot$cm, but can become sizeable in the next generation experiments aiming at $d_n < 10^{27}$ e$\cdot$cm \quad \cite{chupp2019electric}. With the neutron storage cell of height 12 cm, the geometric magnetic field induced spread of the false EDM within the ensemble of stored neutrons can be as large as
\begin{equation}
\Delta d_{false} = \pm \frac{h{\bf \omega}_z}{c}\mu_n \simeq \pm 5\times 10^{-27} {\text e}\cdot\text{cm} \, . \label{eq:dFalseSpread}
\end{equation}

\subsection{Geometric  magnetic field in the all electric proton  EDM storage rings}

\subsubsection{Derivation of GMF in the storage ring geometry}

The storage ring is a cylinder capacitor with the gap $d$
which is  much smaller compared to the height of cylinders $h$, which in its turn is  much smaller than radii of cylinders $r_{1,2}=\rho\mp d/2$. In view of $d\ll h\ll\rho$ we neglect the dependence on the vertical coordinate and have the two-dimensional geometry. The beam trajectory is in the midplane of the storage ring at the orbit radius $|{\bf r}|=\rho$. The electric field in the gap is given by
\begin{equation}
{\bf E}_0=-{\cal E}_0\frac{\rho{\bf r}}{r^2} = -\nabla A_{0}(r), \quad\quad A_0(r) = {\cal E}_0 \rho \ln\frac{r}{\rho}\, .
\label{IIE10}
\end{equation}

It is instructive to start with the storage ring located on the North or South pole. From the viewpoint of distant observer in the reference frame $K'$, the ring rotates with the Earth's rotation angular velocity ${\boldsymbol \omega}$ and linear velocity ${\bf v}({\bf r}) =[{\boldsymbol \omega}\times {\bf r}]$. The static charges on the two rotating cylinders produce the opposite currents and generate in the gap the magnetic fields of the same sign and magnitude. The net result is the magnetic field
\begin{equation}
{\bf H}'_{\boldsymbol\omega}({\bf r})=\frac{1}{c}\big[{\bf v}({\bf r})\times{\bf E}_0({\bf r})\big].
\label{IIE20}
\end{equation}

However, it is basically the motional magnetic field and, to the experimenter in the polar laboratory  it vanishes entirely.
But such an exact cancellation does not hold at an  arbitrary  latitude. In the generic case, the result  (\ref{IIE20}) for the magnetic field  suggests the small parameter $\eta_{\boldsymbol\omega}={|{\boldsymbol\omega}|\rho}/{c}$, similar to that appearing in Eq. (\ref{eq:dFalse}). For the storage ring of radius $\rho \sim 40\mbox{m}$ we have
\begin{equation}
\eta_{\boldsymbol\omega}=\frac{|{\boldsymbol\omega}|\rho}{c}\sim 10^{-11}.
\label{IIE30}
\end{equation}
which is four orders in magnitude larger than the target value $\eta_{EDM}\sim 10^{-15}$.

The electric field (\ref{IIE10}) suggests  for the magnetic potential $\psi$ the Ansatz
\begin{equation}
\psi=f(r)\cdot({\boldsymbol\omega_t}{\bf r})\, ,
\label{IIE40}
\end{equation}
where ${\boldsymbol\omega}_t$ is a projection of the Earth's angular velocity onto the ring plane. A generic solution to Eq. (\ref{H140}) is
\begin{equation}
f(r)=-\frac{{\cal E}_0\rho}{c}\left(\ln\frac{r}{\rho}+\zeta\right) = -\frac{A_o(r)}{c} - \frac{{\cal E}_0\rho}{c}\zeta,
\label{IIE70}
\end{equation}
and
\begin{equation}
{\bf H}_{\boldsymbol\omega}^i= \frac{2{\boldsymbol\omega}_t^j}{c}A_{ij}({\bf r}) =\Bigg\{\frac{A_0}{c}\delta_{ij}
+\frac{{\cal E}_0\rho}{c}\bigg[\left(\zeta-1/2\right)\delta_{ij}
+\frac12(\delta_{ij}-2{\bf n}_i{\bf n}_j)\bigg]\Bigg\}
{\boldsymbol \omega}_t^j,
\label{IIE80}
\end{equation}
where ${\bf n}= {\bf r}/r$ .

The constant $\zeta$ is fixed by the boundary condition that the electric potential $A_{0}$ vanishes rapidly beyond the capacitor, so that in the integral representation for $\psi$ one can perform the integration by parts:
\begin{equation}
\psi({\bf r})=\frac{2{\boldsymbol\omega}_t^j}{c}\int\d^{(2)}y\,\Delta^{-1}({\bf r}-{\bf y}){\bf E}^j_0({\bf y})
\nonumber \\
=-\frac{2{\boldsymbol\omega}_t^j}{c}\int\d^{(2)}y\,\partial_j\Delta^{-1}({\bf r}-{\bf y})A_0(\bf y),
\label{IIE90}
\end{equation}
where $\Delta^{-1}({\bf r})=(1/2\pi)\ln|{\bf r}|$ is the inverse to the Laplace operator. The resulting equation for the symmetric matrix $A_{ij}({\bf r})$
\begin{equation}
A_{ij}({\bf r})=\int\d^{(2)}y\,\partial_i\partial_j\Delta^{-1}({\bf r}-{\bf y})A_0({\bf y}).
\label{IIE100}
\end{equation}
entails
\begin{equation}
\tr A({\bf r})=A_0({\bf r}).
\label{IIE110}
\end{equation}
Hence the expansion of $A_{ij}$ into irreducible tensor structures is of the form
\begin{equation}
A_{ij}({\bf r})=\frac{1}{2}\delta_{ij}A_0({\bf r}) + \sigma({\bf r})(\delta_{ij}-
2{\bf n}^i{\bf n}^j) \, ,
\label{IIE120}
\end{equation}
and a comparison to (\ref{IIE80}) gives immediately
\begin{equation}
\sigma({\bf r}) = \frac{1}{4}{\cal E}_0\rho, \quad \zeta = \frac{1}{2}.
\label{IIE130}
\end{equation}
Our final result for the geometric magnetic field in the gap of the storage ring is
\begin{equation}
{\bf H}_{\boldsymbol\omega}^i=\frac{{\cal E}_0\rho}{c}\Bigg\{\ln\left(\frac{r}{\rho}\right)\cdot\delta_{ij}+
\left(\frac12\delta_{ij}-{\bf n}_i{\bf n}_j\right)\Bigg\}
{\boldsymbol \omega}_t^j\simeq  \frac{{\cal E}_0\rho}{2c}
\left(\delta_{ij}-2{\bf n}_i{\bf n}_j\right){\boldsymbol \omega}_t^j,
\label{IIE1300}
\end{equation}
where in the last step we neglected $|\log(r/\rho)| < d/(2\rho) \ll 1$.

\subsubsection{The background EDM rotations in the geometric and Earth's magnetic fields}

Important virtue of all electric rings is a cancellation of many systematic effects when one compares spin rotations of simultaneously stored CW  and ACW rotating protons. The background magnetic fields spoil the identity of CW and ACW orbits, for a detailed  discussion see the recent monographic document by the CPEDM (Charged Particles EDM) collaboration \cite{abusaif2019storage}. The principal idea behind the all electric rings, run at the so-called magic energy, is to eliminate the motional magnetic field acting on the proton MDM. To this end, on the same footing as GMF ${\bf H_\omega}$ comes the Earth's magnetic field ${\bf H_\oplus}$. The suppression of the latter by magnetic screening can not be ideal one  (see \cite{abusaif2019storage}, Appendix B).

The radial magnetic fields are the most dangerous ones, see Section 2.5.
To this end,  Earth's magnetic field ${\bf H}_\oplus$ and the GMF ${\bf H}_{\bf \omega}$ do differ markedly. In the ring plane, ${\bf H}_{\bf \omega}$ is a uniform one pointing along the meridian ${\bf H}_\oplus ^t = (0, H_\oplus ^t )$. In contrast to that, the geometric magnetic field has the quadrupole-like behaviour along the particle orbit, ${\bf H}_{\bf \omega} = H_{\bf \omega}(\sin 2\theta, \cos 2\theta) $. Here the angular position of the particle in a ring,  $\theta$, is defined by ${\bf n} = (\cos \theta, -\sin \theta)$. In the comoving frame the two radial fields  are  equal to  $H_\oplus^{(r)} = ({\bf n}\cdot {\bf H}_\oplus) = -H_\oplus \sin \theta $  and $H_{\bf \omega}^{(r)} = ({\bf n}\cdot {\bf H}_{\bf \omega}) = H_{\bf \omega} \sin \theta $.
The frozen spin ring operates at the integer spin tune resonance, {i.e.,} in the spin transparent mode \cite{Senichev,Filatov}. In this regime, to the first order of perturbation theory, the spin rotation per turn is proportional to the Bogoliubov-Krylov-Mitropolsky (BKM) averaged \cite{BKM} magnetic fields in the comoving frame,
\begin{equation}
\langle H_\oplus^{(r)} \rangle = \frac{1}{2\pi} \oint d\theta H_\oplus^{(r)}  = \langle H_\omega^{(r)} \rangle =\frac{1}{2\pi}\oint d\theta H_{\bf \omega}^{(r)} =0\, . \label{eq:EDMlike}
\end{equation}
We conclude that to the linear approximation these magnetic fields do not produce the false EDM signal (see also \cite{abusaif2019storage}, Appendix B).

To be on the safe side,  one needs a further dedicated analysis of the false spin rotations with simultaneous allowance for the orbit distortions by the above two background fields. Specifically, one must be aware of a possible geometric spin phase (see also \cite{abusaif2019storage}, Appendix B.2)  Furthermore, one needs to pay an attention to  a possible cross talk between the impact of the geometric magnetic field and the residual Earth's magnetic field. It is an important complex issue on its own to be addressed to in the future.

\subsection{Non-magnetic background to the EDM from the Earth rotation}

Here we evaluate the practical impact on searches for the EDM of the three projections of
${\boldsymbol\Omega}^{GR}_{\boldsymbol\omega}$:
\begin{equation}
{\boldsymbol \Omega}^{GR}_{\boldsymbol\omega} = \frac{1}{\beta} \Omega^{GR}_{{\boldsymbol\omega},r} [{\boldsymbol \beta}\times{\bf n}_{\oplus}] +  \frac{1}{\beta}\Omega^{GR}_{{\boldsymbol\omega},\beta} {\boldsymbol \beta} + \Omega^{GR}_{{\boldsymbol\omega},n_{\oplus}} {\bf n}_{\oplus}.
\nonumber
\end{equation}
We make manifest use of the closed orbit conditions, $({\boldsymbol \beta} \cdot \bf{n}_\oplus) =0,\, ({\boldsymbol n} \cdot \bf{n}_\oplus) =0,\, [{\boldsymbol \omega} \times \bf{n}_\oplus] = [{\boldsymbol \omega}_t \times \bf{n}_\oplus],\, ({\boldsymbol \omega}\cdot  {\boldsymbol \beta})= ({\boldsymbol \omega}_t\cdot  {\boldsymbol \beta})  \,$.

To begin with wee note that the term
\begin{equation}
-\frac{1-2(\gamma^2-1)G}{\gamma}
\left(1-\frac{r_g}{2R_{\oplus}}(1-I)\right)
{\boldsymbol\omega} \label{ConstantOmega}
\end{equation}
in ${\boldsymbol\Omega}^{GR}_{\boldsymbol\omega}$ can be dubbed a kinematic contribution from the Earth's rotation. In its action on spin, it plays exactly the same
role as the Earth's uniform magnetic field $\bf{H}_\oplus$.

We consider first the fake EDM signal from the radial projection
\begin{gather}
\Omega^{GR}_{\omega,r} \equiv \frac{1}{\beta}\left([{\boldsymbol\beta}\times{\bf n}_{\oplus}]
{\boldsymbol{\Omega}}^{GR}_{\omega}\right)
\nonumber \\
=\frac{1-2G(\gamma^2-1)}{\gamma\beta}
\left(1-\frac{r_g}{2R_{\oplus}}(1-I)\right)\big({\boldsymbol\beta} [{\boldsymbol\omega}_t,\,{\bf n}_{\oplus}]\big)
\nonumber \\
-\frac{1}{2\gamma\beta}\Big(\left(5\gamma^2-3\right)G-3\Big)\frac{R_{\oplus}}{c}\big({\boldsymbol\beta}[{\boldsymbol\omega}_t,\,{\bf n}_{\oplus}]\big)^2\nonumber\\
-\gamma G\frac{R_{\oplus}}{c \beta}({\boldsymbol\omega}_t{\boldsymbol\beta})^2\, .
\label{RC70}
\end{gather}
Locally, the first term has a magnitude of the order of the angular velocity of Earth's rotation, $\omega = 7.3\cdot 10^{-5}\mbox{sec}^{-1}$.
At extremely small $\eta_{EDM}\sim 10^{-15}$ of interest, it exceeds by five orders of magnitude the angular velocity of the EDM-induced spin rotation. However, precisely as was the case with the radial projection of the  Earth's magnetic field  $\bf{H}_\oplus$, its effect on the EDM  vanishes upon the BKM averaging \cite{VergelesJETP}. The scale for the  BKM averaged quadratic terms is set by
\begin{equation}
\frac{R_\oplus \omega}{c}\omega \approx 10^{-10} \mbox{rad\ s}^{-1}  ,
\label{eq:omega2}
\end{equation}
which is only one order of magnitude smaller than the EDM signal at $\eta_{EDM}\sim 10^{-15}$.

 Spin rotation about the velocity
${\boldsymbol \beta}$  would gave hindered the EDM driven buildup of the vertical polarization. However, simple considerations show that the BKM averaged spin rotation about the velocity ${\boldsymbol\beta}$ vanishes, $\langle {\boldsymbol\Omega}^{GR}_{\omega,\beta}\rangle =0$.

Spin rotations about   the normal $\boldsymbol{n}_\oplus$ to the storage ring plane spoil the frozen spin condition. After the BKM averaging, we are left with
\begin{equation}
\Omega^{GR}_{{\boldsymbol\omega},n_\oplus}= \
=\frac{2(\gamma^2-1)G-1}{\gamma}
\left(1+\frac{r_g}{2R_{\oplus}}(1-2I)\right)({\boldsymbol\omega}_\cdot{\bf n}_{\oplus})\, ,
\label{RC90}
\end{equation}
which mimics the effect of the vertical magnetic field. As such, it complements a similar contribution from the vertical component of Earth's magnetic field. From the point of view of the proton EDM hunter, the best strategy is the active compensation of these parasitic spin rotations by the comagnetometry technique. Specifically, the buildup of the sideways polarization can be detected by the beam polarimetry  and compensated for applying the vertical magnetic field (a discussion of the multiple bunch comagnetometry is found in Section 10 and Appendix B of Ref. \cite{abusaif2019storage})

To summarize this discussion, to the considered approximation, only the Earth gravity pull, the geometric magnetic field, Earth's magnetic field,  and the kinematic effect from Earth's rotation emerge as serious contenders for systematic background to the EDM signal. What we discussed here is but a tip of iceberg, a more detailed analysis of spin-orbit coupling with allowance for GR effects is called upon.

\section{Conclusions}

We reviewed leading General Relativity effects in spin dynamics of interest for searches of the EDM of neutrons, protons and light nuclei.
Apart from the Earth gravity pull effect with its Standard Candle capacity, of particular interest is the geometric magnetic field in the pure electrostatic systems at rest on the rotating bodies (Earth). From the general relativity point of view, GMF originates from the nonvanishing off-diagonal elements $g_{0i}$ of the metric tensor which are proportional to the angular velocity of rotation of the gravitating body.  In the configuration of experimental setups used in the terrestrial searches for the EDM of neutrons, the geometric magnetic field changes the sign when the electric field is flipped. Consequently, its interaction with the magnetic moment of the neutron can imitate the neutron EDM and that can become a sizeable background in the next generation of the neutron EDM experiments. We found a fairly large background geometric magnetic field in all electric magic storage rings considered a Holy Grail machine for searches of the proton EDM. The symmetry properties  of the geometric magnetic field suggest strong cancellations of its contribution to the proton spin rotations. Still, its impact on the signal of EDM remains an open issue - here one needs a dedicated analysis with full allowance for the spin-orbit dynamics in the storage ring.

This contribution is a humble tribute to the memory of Vladimir Naumovich Gribov, great man of science and the head of the High Energy Physics Department of the Landau Institute, both authors had a privilege to belong to. We are grateful to Julia Nyiri for an invitation to contribute to this volume.
\section*{Acknowledgements}


N.N.N. acknowledges a support by the Russian Fund for Basic
Research (Grant No. 18-02-40092MEGA). The work of S.N.V. was supported by the Ministry of Science and Higher Education of the Russian Federation (state program No. 0033-2019-0005).

\appendix

\section{Geometry.}


We consider stationary metric in the laboratory  reference frame K. Following the Landau-Lifshitz textbook \cite{landau2013classical}, we diagonalize  the quadratic form
\begin{equation}
\d s^2=g_{\mu\nu}\d x^{\mu}\d x^{\nu}
=g_{00}\left(\d x^0+\frac{g_{0i}\d x^i}{g_{00}}\right)^2
-\left(-g_{ij}+\frac{g_{0i}g_{0j}}{g_{00}}\right)\d x^i\d x^j
=\eta_{ab}\left(e^a_{\mu}\d x^{\mu}\right)
\left(e^b_{\nu}\d x^{\nu}\right),
\label{G10}
\end{equation}
where the field $e^a_{\mu}(x)$ is called tetrad,
$a,b,\ldots=0,1,2,3,\, a=(0,\alpha), \,\alpha=1,2,3$, 
\linebreak
$\eta_{ab}=\diag(1,-1,-1,-1)$.
The choice of tetrad is akin to the choice of gauge and is a matter of convenience. The above Landau-Lifshitz choice is the best suited one for description of spin physics on rotating Earth, see Appendix B.
Two  infinitesimally close events are simultaneous if 1-form
\begin{equation}
e^0_{\mu}\d x^{\mu}=\sqrt{g_{00}}\left(\d x^0+\frac{g_{0i}\d x^i}{g_{00}}\right)=0.
\label{G20}
\end{equation}
Then,  the squared interval between simultaneous events  is
\begin{equation}
-\d s^2=\left(-g_{ij}+\frac{g_{0i}g_{0j}}{g_{00}}\right)\d x^i\d x^j
=\sum_{\alpha=1}^3\left(e^{\alpha}_i\d x^i\right)\left(e^{\alpha}_j\d x^j\right),
\label{G30}
\end{equation}
what defines the spatial metric
\begin{equation}
{\mg}_{ij}=\left(-g_{ij}+\frac{g_{0i}g_{0j}}{g_{00}}\right).
\label{G35}
\end{equation}

The local orthonormal basis (ONB) $\tilde{e}^{\mu}_a(x)$ is defined by the equations
\begin{equation}
e^a_{\mu}(x)\tilde{e}_b^{\mu}(x)=\delta^a_b, \quad g_{\mu\nu}\tilde{e}_a^{\mu}\tilde{e}_b^{\nu}=\eta_{ab}.
\label{G40}
\end{equation}
Since according to (\ref{G30})
\begin{equation}
e^{\alpha}_0=0,
\label{G50}
\end{equation}
one readily finds:
\begin{equation}
\tilde{e}^i_0=0, \quad  \tilde{e}^0_0=\left(e^0_0\right)^{-1},
\quad
\tilde{e}^i_{\alpha}e_i^{\beta}=\delta^{\beta}_{\alpha},
\quad \tilde{e}^0_{\alpha}=-\left(e^0_0\right)^{-1}e^0_i\tilde{e}^i_{\alpha}.
\label{G60}
\end{equation}

The   rules of the tensor component transformation from the coordinate basis to the ONB  and vice versa are  standard ones.
For example
\begin{equation}
X^a=e^a_{\mu}X^{\mu}, \quad X^{\mu}=\tilde{e}^{\mu}_aX^a, \quad
\xi_a=\tilde{e}_a^{\mu}\xi_{\mu}.
\label{G64}
\end{equation}
In ONB the tensor indices are lowered and raised with the help of metric tensors $\eta_{ab}$ and $\eta^{ab}$.
With the above chosen tetrad there is a complete equivalence between  $J^i=0$ and $J^{\alpha}=0$:
\begin{equation}
J^a=e^a_{\mu}J^{\mu}=\left(e^0_0J^0,\,e^{\alpha}_iJ^i\right)=\left(e^0_0J^0,\,0,0,0\right).
\nonumber
\end{equation}

The covariant derivatives $\nabla_{\mu}$ in the coordinate
basis and in ONB are related as
\begin{gather}
\nabla_{\mu}X^a\equiv e^a_{\nu}\nabla_{\mu}X^{\nu}=\partial_{\mu}X^a+\gamma^a_{b\mu}X^b,
\nonumber \\
\nabla_cX^a=\tilde{e}^{\mu}_c\nabla_{\mu}X^a=\tilde{e}^{\mu}_c\partial_{\mu}X^a+\gamma^a_{bc}X^b,
\label{G65}
\end{gather}
where $\gamma^a_{bc}\equiv\tilde{e}^{\mu}_c\gamma^a_{b\mu}$ are connection coefficients,
\begin{equation}
\gamma_{abc}\equiv\eta_{ad}\gamma^d_{bc}=-\gamma_{bac}.
\label{G70}
\end{equation}
The condition that connections are free of torsion reads
\begin{equation}
\partial_{\mu}e^a_{\nu}-\partial_{\nu}e^a_{\mu}+\gamma^a_{b\mu}e^b_{\nu}-\gamma^a_{b\nu}e^b_{\mu}=0.
\label{G80}
\end{equation}
Then, connection coefficients are determined uniquely by Eqs. (\ref{G70}) and (\ref{G80}):
\begin{gather}
\gamma_{abc}=\frac12\left(C_{abc}-C_{bac}-C_{cab}\right),
\nonumber \\
C_{abc}\equiv \eta_{ad}C^d_{bc}, \quad C^a_{bc}=(\partial_ie^a_{\nu})\left(\tilde{e}^i_b\tilde{e}^{\nu}_c-\tilde{e}^i_c\tilde{e}^{\nu}_b\right).
\label{G90}
\end{gather}
The last equality in (\ref{G90}) is valid only in the case of a time-independent metric of our interest.

Under the local Lorentz transformation
\begin{equation}
\tilde{e}'^{\mu}_a(x)=\Lambda^b_a(x)\tilde{e}^{\mu}_b(x)
\label{G91}
\end{equation}
the connection coefficients transform as
\begin{equation}
\gamma'^{\,a}_{b\,\mu}=(\Lambda^{-1})^a_c\Lambda^d_b\gamma^c_{d\,\mu}+(\Lambda^{-1})^a_c\partial_{\mu} \Lambda^c_b.
\label{G92}
\end{equation}
Here  $\Lambda^a_b(x)$ is a local  Lorentz transformation matrix.

\section{Metric, tetrad and connection coefficients}


The rotating reference frame K is  defined for the  Earth rotating with constant angular velocity ${\boldsymbol\omega}$ in the inertial frame  of distant stars $K'$. The local coordinates,
vectors etc. in $K'$ are denoted as $x^{\prime\mu}$, ${\bf R}'$ and so forth. The space-time of $K'$ is described by the Kerr metric of the rotating  Earth \cite{landau2013classical}. For the purposes of our analysis it is sufficient to use a limit of weak gravity  and nonrelativistic rotation velocity and we expand the Kerr metric retaining the terms linear in ${\boldsymbol{\omega}}$ and $r_g$, bilinear in $r_g$ and ${\boldsymbol{\omega}}$
and  quadratic in ${\boldsymbol{\omega}}$:
\begin{gather}
g'_{00}({\bf R}')=1-\frac{r_g}{R'},
\nonumber \\ g'_{0i}({\bf R}')=\frac{2kC}{c^3R'^3}\big[{\boldsymbol\omega}\times {\bf R}'\big]^{i}=I\cdot\frac{r_gR_{\oplus}^2}{R^{\prime3}}\frac{\big[{\boldsymbol\omega}\times {\bf R}'\big]^{i}}{c},
\nonumber \\
g'_{ij}({\bf R}')=-\left(1+\frac{r_g}{R'}\right)\delta_{ij},
\nonumber \\
r_g=\frac{2kM_{\oplus}}{c^2},
\label{MetStar10}
\end{gather}
where $R'=|{\bf R}'|$, $k=6,674\cdot10^{-8}\mbox{cm}^3\cdot\mbox{g}^{-1}\cdot\mbox{sec}^{-2}$ is the gravitation constant, $M_{\oplus}$ and  $C=IM_{\oplus}R_{\oplus}^2$ are the Earth mass  and  moment of inertia relative to polar axis, $I=0.3307$,
$M_{\oplus}=5,972\cdot10^{27}\mbox{g}$, $R_{\oplus}=6,378\cdot10^8\mbox{cm}$,
$r_g=2kM_{\oplus}/c^2=0,887\mbox{cm}$.  Next we transform the metric (\ref{MetStar10}) into the metric in the laboratory frame $K$. We take the coordinates in $K$ and $K'$
having the same origin at the centre of Earth.
The local coordinates in the frame $K$ are denoted as $x^{\mu}$ and, by definition, they are connected with coordinates $x^{\prime\mu}$
as follows,
\begin{equation}
\d x^{\prime0}=\d x^0, \quad \d {\bf R}'=\d {\bf R}+[{\boldsymbol{\omega}}\times {\bf R}]\frac{\d x^0}{c},
\quad |{\bf R}'|=|{\bf R}|\, .
\label{CoordSubst10}
\end{equation}
The metric $g_{\mu\nu}$
in the frame $K$ equals
\begin{gather}
g_{00}=1-\frac{r_g}{R}-\frac{[{\boldsymbol\omega}\times {\bf R}]^2}{c^2},
\nonumber \\
g_{0i}=-\Bigg\{1+\frac{r_g}{R}\left(1-I\cdot\frac{R_{\oplus}^2}{R^2}\right)\Bigg\}\frac{[{\boldsymbol\omega}\times{\bf R}]^i}{c},
\nonumber \\
g_{ij}=-\left(1+\frac{r_g}{R}\right)\delta^{ij}.
\label{metric20}
\end{gather}
The proper time $t$ in the laboratory frame is related to  $ x^0$ in
Eq. (\ref{CoordSubst10}) as
\begin{equation}
\d t=\frac{1}{c}\sqrt{g_{00}}\d x^0.
\label{PT}
\end{equation}
The inverse metric tensor $g^{\mu\nu}=\eta^{ab}\tilde{e}^{\mu}_a\tilde{e}^{\nu}_b$ is
\begin{gather}
g^{00}=1+\frac{r_g}{R},
\nonumber \\
g^{0i}=-\Bigg\{1+\frac{r_g}{R}\left(1-I\cdot\frac{R_{\oplus}^2}{R^2}\right)\Bigg\}\frac{[{\boldsymbol\omega}\times {\bf R}]^i}{c},
\nonumber \\
g^{ij}=-\left(1-\frac{r_g}{R}\right)\delta^{ij}+
\frac{[{\boldsymbol\omega}\times {\bf R}]^i[{\boldsymbol\omega}\times{\bf R}]^j}{c^2}.
\label{IM}
\end{gather}
To the same approximation the tetrad equals
\begin{gather}
e^0_0=\sqrt{g_{00}}=1-\frac{r_g}{2R}-\frac{[{\boldsymbol\omega}\times {\bf R}]^2}{2c^2}, \quad e^{\alpha}_0=0,
\nonumber \\
e^0_i=\frac{g_{0i}}{\sqrt{g_{00}}}=-\Bigg\{1+\frac{r_g}{R}\left(\frac32-I\cdot\frac{R_{\oplus}^2}{R^2}\right)\Bigg\}
\frac{[{\boldsymbol\omega}\times{\bf R}]^i}{c},
\nonumber \\
e^{\alpha}_i=\left(1+\frac{r_g}{2R}\right)\delta^{\alpha}_i+
\frac{[{\boldsymbol\omega}\times {\bf R}]^{\alpha}[{\boldsymbol\omega}\times {\bf R}]^i}{2c^2}\, ,
\label{tetrad}
\end{gather}
and the ONB vector fields are
\begin{gather}
\tilde{e}_0^0=1+\frac{r_g}{2R}+\frac{[{\boldsymbol\omega}\times {\bf R}]^2}{2c^2}, \quad   \tilde{e}_0^i=0,
\nonumber \\
\tilde{e}_{\alpha}^0=\Bigg\{1+\frac{r_g}{R}\left(\frac32-I\cdot\frac{R_{\oplus}^2}{R^2}\right)\Bigg\}
\frac{[{\boldsymbol\omega}\times {\bf R}]^{\alpha}}{c},
\nonumber \\
\tilde{e}^i_{\alpha}=\left(1-\frac{r_g}{2R}\right)\delta_{\alpha}^i-
\frac{[{\boldsymbol\omega}\times {\bf R}]^i[{\boldsymbol\omega}\times {\bf R}]^{\alpha}}{2c^2}.
\label{ONB}
\end{gather}
Here we note that
\begin{equation}
\tilde{e}^i_{\alpha}e_j^{\alpha}=\delta^i_j, \quad
{\mg}_{ij}\tilde{e}^i_{\alpha}\tilde{e}^j_{\beta}=\delta_{\alpha\beta}.
\label{D40}
\end{equation}
 and the set $\{\tilde{e}^i_{\alpha}\}$ plays the role of a purely spatial ONB in the purely spatial metric (\ref{metric20}).

 Now we turn to the definition (\ref{G90}) and  report the connection coefficients in our ONB (\ref{ONB}):
\begin{gather}
\gamma_{\alpha 00}=(1/c^2)\left({\bf g}_0-\big[{\boldsymbol\omega}\times [{\boldsymbol\omega}\times {\bf R}]\big]\right)^{\alpha},
\nonumber \\
\gamma_{\alpha 0\beta}=\gamma_{\alpha\beta 0}
\nonumber \\
=\varepsilon_{\alpha\beta\sigma}\bigg\{\left(1-\frac{r_g}{2R_{\oplus}}(1-I)\right)
\frac{{\boldsymbol\omega}^{\sigma}}{c}
+\frac{r_g}{R_{\oplus}}\left(1-\frac32I\right)\frac{({\boldsymbol\omega}{\bf n}_{\oplus}){\bf n}_{\oplus}^{\sigma}}{c}\bigg\},
\nonumber \\
\gamma_{\alpha\beta\rho}=\varepsilon_{\alpha\beta\sigma}\bigg\{-
\varepsilon_{\sigma\rho\delta}\frac{{\bf g}_0^{\delta}}{c^2}
+\frac{3}{2c^2}{\boldsymbol\omega}^{\sigma}[{\boldsymbol\omega}\times{\bf R}]^{\rho}\bigg\}.
\label{Connection}
\end{gather}


It is appropriate to  present here a technical observation why the Landau-Lifshitz tetrad is custom tailored for description of the cyclic motion of a particle in a storage ring, periodically probing one and the same gravitational field. While in the 4D space the world line of particle is a helix, in the 3D laboratory space $x^i(s)$ the particle periodically rotates on the closed circular orbit,  $x^i(s_0+cT)=x^i(s_0)=x^i_0$, where T is the revolution period.  The same  periodicity is exhibited by
the 4-velocity $u^\alpha$, with  $u^0$ preserved during rotation in the storage ring. The latter statement presumes that the gravity potential relief is constant on the scale of the storage ring $\rho$, which is a small parameter in the problem. More technical formulation of this smallness will be presented below.

Specifically, of our interest is the parallel transport of $\bar{u}^\alpha$ per turn, defined by the closed 4D loop integral
\begin{equation}
\oint_{\cal L} \overline{u}^a(x(s))\d s\,,
\label{torsion}
\end{equation}
where
\begin{equation}
\overline{u}^{\alpha}(x(s))=\overline{e}^{\alpha}_{\mu}(x(s))\frac{\d x^{\mu}(s)}{\d s}\, ,
\label{paralltransf}
\end{equation}
describes the parallel transport of vector $u^{\alpha}(x(s))$ from the running points $x^{\nu}(s)$
to the initial point $x^{\nu}_0$ along loop ${\cal L}$. This loop in the 4D space can be split into two open contours: ${\cal L}_1$ describes the helical motion $ x^{\mu}(s) = (s,{\vec x}(s)), \quad
s_0\leq s \leq s_0+cT$ subject to  the storage ring periodicity condition,
${\vec x}(s_0+cT)= {\vec x}(s_0) ={\vec x}_0$, while the contour ${\cal L}_2$ closes the 4D loop ${\cal L}$:
$x^{\mu}(s) = (s,{\vec x}_0),\quad
s_0+cT\geq s\geq s_0$.
In the above definition of the parallel transport,
\begin{gather}
\overline{e}^a_{\mu}(x(s))=e^a_{\mu}(x(s))+\delta e^a_{\mu}(x(s)),
\nonumber \\
\delta e^a_{\mu}(x(s))=-(\gamma^a_{bc}e^b_{\mu}e^c_{\nu})\Big|_{x^{\mu}_0}
\big(x^{\nu}(x(s))-x^{\nu}_0\big)
\label{paralltransf_e}
\end{gather}
the gravity effects on the storage ring scale $\rho\sim10^4\,\mbox{cm}$  can be  treated to the linear approximation.

As well known, the infinitesimal loop integral
$$I^a=\oint_{\cal L} \overline{u}^a(x(s))\d s$$ is proportional to the torsion tensor
and bivector $\sigma^{\mu\nu}=\int_{\sigma}\d x^{\mu}\wedge\d x^{\nu}$, where $\sigma$ is a surface spanning  the
boundary {\cal L}. The bivector $\sigma^{\mu\nu}$ depends on  the loop {\cal L} but not on  the shape of the spanned surface  $\sigma$.
Considered 4D spaces are free of torsion and to the linear approximation in $\sigma^{\mu\nu}$ the  above loop integral is equal to zero. To the next order in $|\sigma^{\mu\nu}|$, contributions to $I^a$ proportional, for instance, to
\begin{equation}
\nabla_{\tau}\mR^a_{\nu\lambda\mu}\sigma^{\tau\nu}\sigma^{\lambda\mu}\sim(r_g/R_{\oplus}^4)\rho^4\sim10^{-19}\mbox{cm}
\end{equation}
are possible. To the desired accuracy, they are entirely negligible, and for   $a\longrightarrow\alpha$ we can put
\begin{equation}
I^a= \int_{{\cal L}_1}\overline{e}^{\alpha}_{\mu}(s))\d x^{\mu}(s)
-\int_{s_0}^{s_0+cT}\overline{e}^{\alpha}_0(s_0+s)\d s=0.
\label{zerotorsion}
\end{equation}
With reference to Eq. (\ref{D40}) we can invert the definition  (\ref{D22N}):
\begin{equation}
\frac{\d x^i}{\d s}=\tilde{e}^i_{\alpha}u^{\alpha},
\label{D30}
\end{equation}
so that in this ONB $(\d x^i/\d s)$ is a vector of purely spatial velocity. Then, according to (\ref{paralltransf_e}),    $e^{\alpha}_0=0$ entails $\overline{e}^{\alpha}_0=0$.
Therefore, the second integral in Eq.  (\ref{zerotorsion}) just vanishes, while the first one simplifies to:
\begin{equation}
\oint_{{\bar {\cal  L}}_1}(\overline{e}^{\alpha}_i(s))\d x^i(s)=0.
\label{D110}
\end{equation}
Here the loop ${\bar {\cal  L}}$ is defined in the 3D hyperspace at $x^0 = {\rm const.}$ Eq. (\ref{D110}) can be taken for the mathematical confirmation that the ONB based on the  Landau-Lifshitz tetrad provides an adequate description of the particle motion in cyclic accelerators.

\section{Useful relations}

The following  relations are used  in the   main body of the text:
\begin{gather}
\varepsilon_{abcd}\tilde{e}^{\mu}_a\tilde{e}^{\nu}_b=\frac{1}{\sqrt{-g}}
\varepsilon_{\mu\nu\lambda\rho}e^c_{\lambda}e^d_{\rho},
\nonumber \\
\varepsilon_{abcd}\tilde{e}^{\mu}_a\tilde{e}^{\nu}_b\tilde{e}^{\lambda}_c=\frac{1}{\sqrt{-g}}
\varepsilon_{\mu\nu\lambda\rho}e^d_{\rho},
\nonumber
\end{gather}
which imply that
\begin{gather}
\varepsilon_{\alpha\beta\gamma}\tilde{e}^i_{\alpha}\tilde{e}^j_{\beta}=\frac{1}{\sqrt{-g}}e^0_0
\varepsilon_{ijk}e^{\gamma}_k,
\nonumber \\
\varepsilon_{\alpha\beta\gamma}\tilde{e}^i_{\alpha}\tilde{e}^0_{\beta}=\frac{1}{\sqrt{-g}}
\varepsilon_{ijk}e^{\gamma}_je^0_k,
\nonumber \\
\varepsilon_{ijk}e^{\alpha}_ie^{\beta}_j=\sqrt{-g}\,\varepsilon_{\alpha\beta\gamma}
\tilde{e}^0_0\tilde{e}^k_{\rho},
\nonumber \\
\varepsilon_{\alpha\beta\gamma}\tilde{e}^i_{\alpha}\tilde{e}^0_{\beta}\tilde{e}^j_{\gamma}=\frac{1}{\sqrt{-g}}
\varepsilon_{ijk}e^0_k,
\label{H25n}
\end{gather}
and so forth.

\section{Riemann normal coordinates}

To elucidate the interpretation of EM fields in a curved space-time it is useful to recall dynamic equations in the Riemann normal coordinates. The
Riemann normal coordinates $y^{\mu}$ can be introduced in the vicinity of any point $p$, so that at  the origin
$y^{\mu}(p)=0$ and the small-$y^{\mu}$ expansions read
\begin{gather}
e^a_{\mu}(y)=\delta^a_{\mu}+\frac16\mR^a_{\nu\lambda\mu}(p)y^{\nu}y^{\lambda}+\O(y^3),
\nonumber \\
g_{\mu\nu}(y)=\eta_{\mu\nu}+\frac13\mR_{\mu\lambda\rho\nu}(p)y^{\lambda}y^{\rho}+\O(y^3),
\nonumber \\
\eta_{\mu\nu}=\diag(1,\,-1,\,-1,\,-1),
\nonumber \\
\gamma_{ab\,\mu}(y)=\frac12\mR_{ab\,\nu\mu}(p)y^{\nu}+\O(y^2),
\label{G160}
\end{gather}
where $\mR_{ab\,\nu\mu}$ is the Riemann curvature tensor. Of course, we mean macroscopic scales and treat hadrons as pointlike particles.

A vicinity of the origin of the Riemann normal coordinates is the mathematical model of a "freely falling lift". Since near the Earth surface $|\mR_{ab\,\nu\mu}|\sim r_g/R_{\oplus}^3\sim0,5\cdot10^{-26}\mbox{cm}^{-2}$,
this exceedingly small curvature can be neglected in "falling lifts" of quite a macroscopic size. The dynamics within the "falling lift" takes the same form as  in the Cartesian coordinates in the Minkowski space with GR effects treated in the post-Newtonian approximation. Evidently, in the  Riemann normal coordinates inside the "falling lift", all field components in the ONB $F^{ab}=
e^a_{\mu}e^b_{\nu}F^{\mu\nu}$ coincide with the corresponding field components in the Riemann coordinates.
Upon transition to arbitrary ONB, all tensors are transformed in accordance with the usual Lorentz rules, and a connection determined by ONB appears in the equations of motion.

The  reference frame (\ref{G160}) will be referred to as $K_0$.


%

\end{document}